# Superconductivity below 2.5K in $Nb_{0.25}Bi_2Se_3$ topological insulator single crystal

M. M. Sharma[1,2], P. Rani[1], Lina Sang[3], X.L. Wang[3] and V.P.S. Awana[1,*]

[1]*CSIR-National Physical Laboratory, K.S. Krishnan Marg, New Delhi-110012, India*
[2]*Academy council of scientific and industrial Research, Ghaziabad U.P.-201002, India*
[3]*Institute of superconducting and electronic materials, University of Wollongong, NSW 2522, Australia*

We report crystal growth and below 2.5K superconductivity of $Nb_{0.25}Bi_2Se_3$. These crystals are grown by self flux method. The X-ray diffraction (*XRD*) pattern of as grown crystal flake shows (00*l*) plane (c-orientation) growth. The Rietveld refinement of crushed crystal powder *XRD* (*PXRD*) pattern confirms the phase purity of the studied sample having R-3m space group of rhombohedral crystalline structure. The Raman spectrum of the studied $Nb_{0.25}Bi_2Se_3$ crystal distinctly shows three well defined vibrational modes in terms of $A^1_{1g}$, $Eg^2$, $A^2_{1g}$ at around 72, 129 and 173cm$^{-1}$, which are slightly shifted in comparison to pure $Bi_2Se_3$. Magnetization studies in terms of field cooled (*FC*) and Zero field cooled (*ZFC*) magnetic susceptibility measurements show the diamagnetic transition ($T_c^{onset}$) of the compound at around 2.5K and near saturation of the same below around 2.1K. The isothernal magnetization (*MH*) being taken at 2K, revealed the lower critical field ($H_{c1}$) of around 50Oe and the upper critical field ($H_{c2}$) of 900Oe. It is clear the studied $Nb_{0.25}Bi_2Se_3$ is a bulk superconductor. The superconducting critical parameters thus calculated viz. the coherence length, upper and lower critical fields and superconducting transition temperature for as grown $Nb_{0.25}Bi_2Se_3$ single crystal are reported here.

Key Words: Doped Topological Insulators, Crystal Growth, Structural Details, and Superconductivity

PACS: 74.70.Dd, 74.62.Fj, 74.25.F-

**[*]Corresponding Author**

Dr. V. P. S. Awana: E-mail: awana@nplindia.org
Ph. +91-11-45609357, Fax-+91-11-45609310
Homepage: awanavps.webs.com

## Introduction

Topological insulators (TIs) with extraordinary quantum properties in terms of their highly conducting surface and bulk insulating transport properties deep within due to time reversal symmetry (TRS) protected surface/edge states had attracted the attention of condensed matter physics community at large [1-3]. Besides several other quantum properties, more recently superconductivity could be induced in TIs by application of hydrostatic pressure [4], superconducting proximity effect, and by doping or intercalation of suitable elements like Cu [5,6], Sr [7,8], Nb [9,10], Tl [11], and Pd [12] etc. Understanding of the surface conductivity of TIs had yet been to some extent clear in terms of Dirac point driven electronic structure. The superconductivity of TIs further complicates the situation in terms of the copper pairs being formed by the edge/surface states driven carriers. Superconducting TIs do further host some novel properties besides the known usual superconductivity phenomenon, viz zero biased conduction peak [13], quantized thermal hall conductivity [14], anomalous Josephson effect [15], odd frequency Cooper pairs [16], and Majorana Fermions [17] etc. It is clear that the superconducting TIs are very much a suitable choice for fundamental research related to quantum phenomenon [4-12].

Keeping in view, that $Bi_2Se_3$ is one of the most studied topological insulators, in current short letter, we report the crystal growth and characterization of $Nb_{0.25}Bi_2Se_3$ superconductor. Although scant reports exist on superconducting TIs [4-17], yet it is important to reproduce the new observations in different laboratories. This is main purpose of the short letter concluding that superconductivity exists in $Nb_{0.25}Bi_2Se_3$ below 2.5K with lower and upper critical fields at 2 K of around 50Oe and 900Oe respectively. This is in conformity with earlier results [9, 10].

## Experimental

$Nb_{0.25}Bi_2Se_3$ single crystal was grown by self-flux method in simple automated furnace via solid state reaction route, as reported by some of us recently [18]. The schematic heat treatment of $Nb_{0.25}Bi_2Se_3$ single crystal is shown in left side inset of Fig. 1. One of the important issues is to not only very slow cooling ($1^0$C/hrs) from $950^0$C melt to $600^0$C, but to quench in ice after annealing for 24 hours at $600^0$C. This is necessary to avoid the segregation of Nb as an impurity in the matrix. Details can be followed from ref. 18. *XRD* on mechanically cleaved crystal flake and powder form had been carried out using Rigaku made Mini Flex II X-ray diffractometer having Cu $K_\alpha$ radiation of 1.5418Å wavelength. Raman spectra of as grown $Nb_{0.25}Bi_2Se_3$ single crystal are recorded at room temperature using the Renishaw Raman Spectrometer. Magnetization studies are done on PPMS (Physical Property measurement System) from Quantum Design (*QD*).

**Results and Discussion**

Figure 1 depicts the Rietveld refinement of gently crushed powder *XRD* pattern of the $Nb_{0.25}Bi_2Se_3$. The studied as grown crystal is crystallized in rhombohedral structure having R-3m space group [19]. All the observed diffraction lines are indexed on the XRD pattern. Hardly any un-reacted peak is seen within X-Ray limit. The fitted parameter ($\chi^2$) value is found to be ~7.43, which is reasonably good as far as Reitveld PXRD is concerned. Atom positions obtained from Rietveld refinement of PXRD data are Bi [0, 0, 0.4018(6)] & Se [0, 0, 0.2086(5)] and lattice parameters of the grown $Nb_{0.25}Bi_2Se_3$ single crystal are $a = b =$ 4.1458(3)Å & $c$ = 28.5759(2)Å of $\alpha = \beta = 90^o$ & $\gamma = 120^o$. The VESTA software drawn representative unit cell of the studied $Nb_{0.25}Bi_2Se_3$ single crystal can be seen in ref. 18. The studied crystal morphology as seen from Scanning Electron Microscope (*SEM*) was seen to be laminar type and the stoichiometric formula being calculated from *EDAX* analysis is close to the nominal compositional value, for details see ref. 18. The right hand side inset of Fig. 1 shows the *XRD* pattern of crystal flake of as grown $Nb_{0.25}Bi_2Se_3$ single crystal confirming the crystalline nature of the same growing in single plane i.e. (00*l*) only. The intense peaks are seen for (003n) with n = 1, 2, 3…. etc.

Figure 2 shows the Raman spectrum for both reported [20] pure $Bi_2Se_3$ and the studied $Nb_{0.25}Bi_2Se_3$ crystal, one over the other. Clearly alike $Bi_2Se_3$, the studied $Nb_{0.25}Bi_2Se_3$ crystal also shows the characteristic vibrational modes for these systems i.e. $A^1_{1g}$, $Eg^2$, $A^2_{1g}$. These modes are seen at 70.66, 130.20 and 174.547cm$^{-1}$ for pure $Bi_2Se_3$ and at 70.29, 129.99 and 172.23cm$^{-1}$ for $Nb_{0.25}Bi_2Se_3$ respectively. The Raman Shift peaks are observed to move to lower frequency side in spectra, while the intensity of the peaks found to be decreasing for $A^1_{1g}$ and $A^2_{1g}$ and increasing for $Eg^2$ mode. The observed lower Raman Shift values for studied $Nb_{0.25}Bi_2Se_3$ crystal may be due to intercalation of Nb atoms residing in vander waals gap between two quintuple layers weakening the bonding forces [19-21].

Figure 3 depicts the magnetization versus temperature response of studied $Nb_{0.25}Bi_2Se_3$ crystal under field cooled (FC) and zero field cooled (*ZFC*) conditions under an applied field of 20Oe. A clear diamagnetic transition appears in both *FC* and *ZFC* plots with a critical temperature near 2.5K, confirming the onset of superconductivity in studied $Nb_{0.25}Bi_2Se_3$ crystal. The diamagnetic signal observed at 2.5K saturates at around 2.1K, thus indicating the superconducting transition width to be around 0.4K. Lower inset in Fig. 3 shows *MH* curve for studied $Nb_{0.25}Bi_2Se_3$ crystal at 2, 2.2 and 2.4K i.e., below the critical temperature. The wide open *MH* plots confirm the observation of bulk superconductivity right up to 2.4K. The lower critical field $H_{c1}$ is found to be 50Oe and upper critical field $H_{c2}$ being 900Oe at 2K. The same are 8.25Oe and 400Oe at 2.2K and 7.10Oe and 200Oe at 2.4K. The lower critical field ($H_{c1}$) is defined as the deviation of linear diamagnetic plot towards mixed state and is marked with a straight line in the inset of Fig. 3. The upper critical field ($H_{c2}$) is roughly known to coincide with the irreversibility field of a type-II superconductor isothermal magnetization (*MH*) plot. This is marked on corresponding plot in inset of Fig. 3. It is clear from Fig. 3, inset that the studied $Nb_{0.25}Bi_2Se_3$ crystal is a type-II superconductor. Critical field for the studied $Nb_{0.25}Bi_2Se_3$ crystal as calculated from the formula $H_c = (H_{c1}*H_{c2})^{1/2}$, is found to be 212.13Oe. at 2K Upper critical field at 0K i.e. $H_{c2}(0)$ is measured by using the GL equation given below, where t represent reduced temperature $T/T_c$.

$$H_{c2}(T) = H_{c2}(0)x[\frac{(1-t^2)}{(1+t^2)}]$$

$H_{c2}(0)$ is found to be $4.1*10^3$ Oe i.e. around 0.4Tesla at 2K. Ginzburg-Landau parameter Kappa (κ) is determined by using formula $H_{c2}(0)=\kappa*2^{1/2}*H_c$ and is found to be 13.66 which is far greater than $1/2^{½}$, hence confirming the type-II superconductivity of studied $Nb_{0.25}Bi_2Se_3$ crystal. Coherence length of $Nb_{0.25}Bi_2Se_3$ crystal is determined by using the formula $H_{c2}(0) = \frac{\varphi_0}{2\Pi\xi(0)^2}$ where $\Phi_0$ is flux quanta. The value of flux quanta is $2.0678 \times 10^{-15}$Wb. Thus calculated coherence length $\xi(0)$ of studied $Nb_{0.25}Bi_2Se_3$ crystal is found to be 2.83Å. Further, penetration depth $\lambda(0)$ is determined by kappa parameter using formula κ =

$\lambda(0)/\xi(0)$ which is found to be 38.65Å at 2K. The fundamental superconducting parameters of studied $Nb_{0.25}Bi_2Se_3$ crystal at 2K are given in Table 1.

## Conclusion

In summary, in this the short letter we studied low temperature magnetic properties of $Nb_{0.25}Bi_2Se_3$ single crystal which confirms that the sample is superconducting. Various superconductivity critical parameters viz. critical field, kappa parameter, coherence length and penetration depth are calculated. Studied $Nb_{0.25}Bi_2Se_3$ single crystal is a Type-II superconductor. The phase purity of the grown crystal is confirmed by PXRD. The Raman shift in vibration modes due to Nb intercalation is also seen and briefly discussed.

## Acknowledgements

Authors would like to thanks CSIR-NPL & Institute of superconducting and electronic materials, University of Wollongong, NSW for all experimental observation of data. M.M. Sharma would like to thanks CSIR for research fellowship and AcSIR-Ghaziabad for PhD registration.

Table 1: Structural and Basic Superconducting Parameters for Studied $Nb_{0.25}Bi_2Se_3$ Single Crystal

| ***Physical Property ($Nb_{0.25}Bi_2Se_3$)*** | ***Corresponding values*** |
|---|---|
| Crystal Structure | Rhombohedral ; R-3m space group<br>$\alpha = \beta = 90^o$ & $\gamma = 120^o$ |
| Lattice parameters | $a = b = 4.1458(3)$Å & $c = 28.5759(2)$Å |
| Superconducting critical temperature & Fields | $T_c^{onset} = 2.5K$, $\Delta T_c = 0.4K$<br>$H_{c1} = 50Oe$, $H_{c2} = 900Oe$, $H_c = 212.13Oe$,<br>$H_{c2}(0) = 4100Oe$ |
| Coherence length | $\xi(0) = 2.83$Å |
| Penetration depth | $\lambda(0) = 38.65$Å |
| Ginzburg-Landau parameter Kappa (κ) | κ = 13.66 |
| Raman Mode positions: | $A^1_{1g} = 70.29cm^{-1}$<br>$Eg^2 = 129.99cm^{-1}$<br>$A^2_{1g} = 172.23cm^{-1}$ |

## Figure Captions

**Figure 1**: Rietveld refinement of gently crushed powder *XRD* pattern, left inset the heat treatment and right inset surface *XRD* of the $Nb_{0.25}Bi_2Se_3$ single crystal.

**Figure 2**: Raman spectrum of $Nb_{0.25}Bi_2Se_3$ and $Bi_2Se_3$ single crystals.

**Figure 3**: *FC* and *ZFC* magnetization (*MT*) of $Nb_{0.25}Bi_2Se_3$ single crystal from 1.5 to 5.5K at 20Oe applied field, the inset shows the isothermal magnetization (*MH*) of the same at 2, 2.2 and 2.4K.

Fig. 1

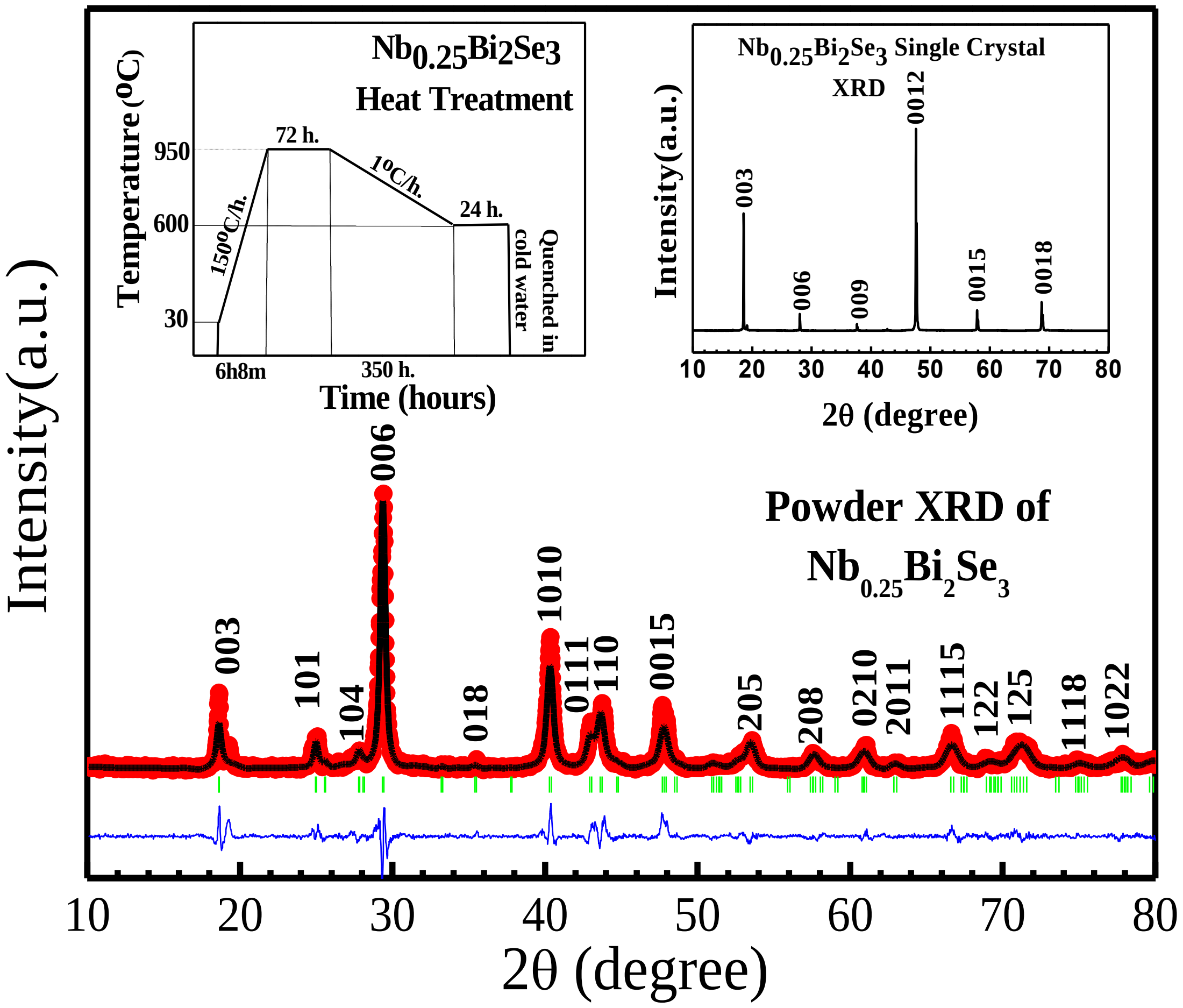

Fig. 2

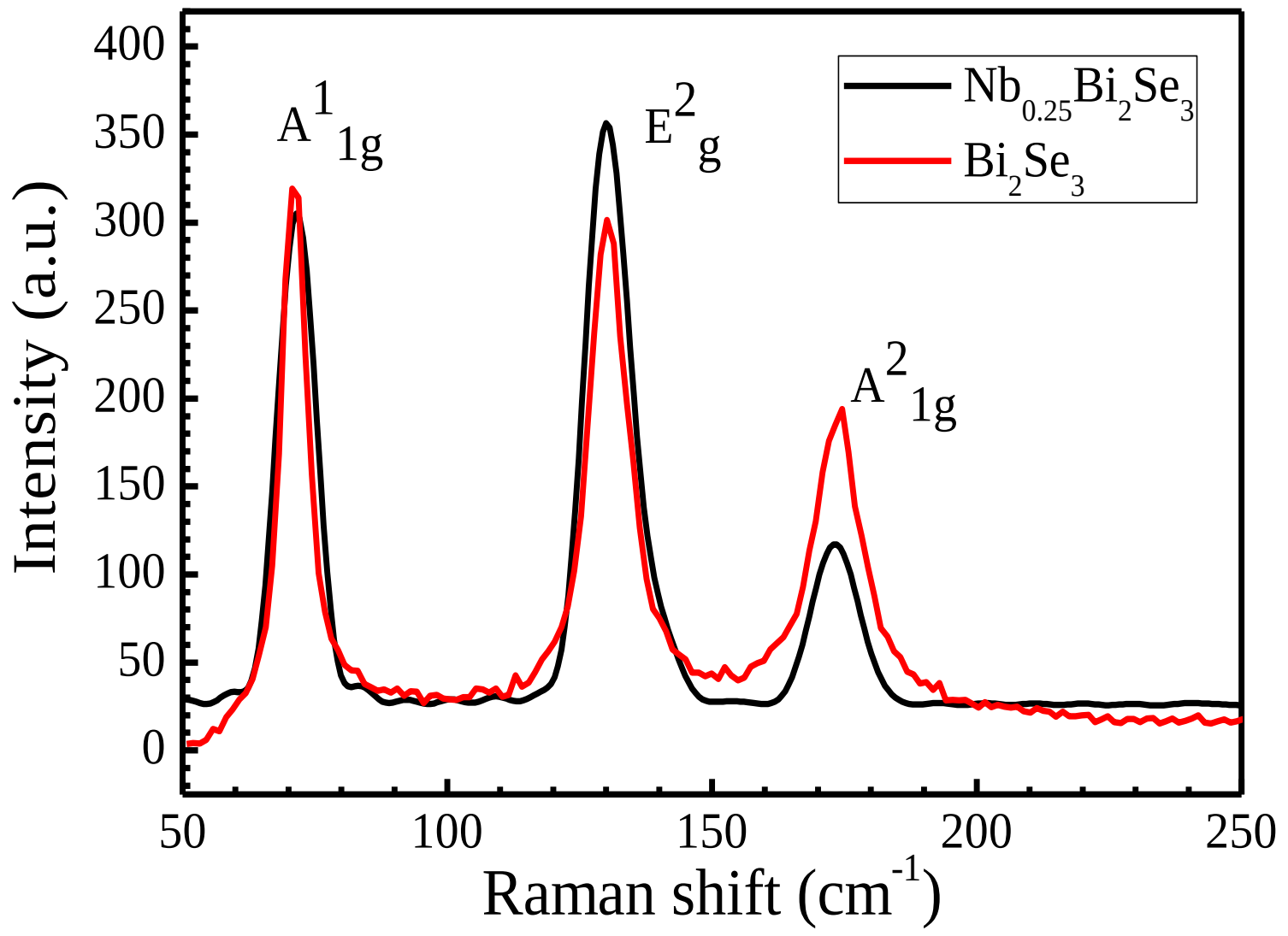

Fig. 3

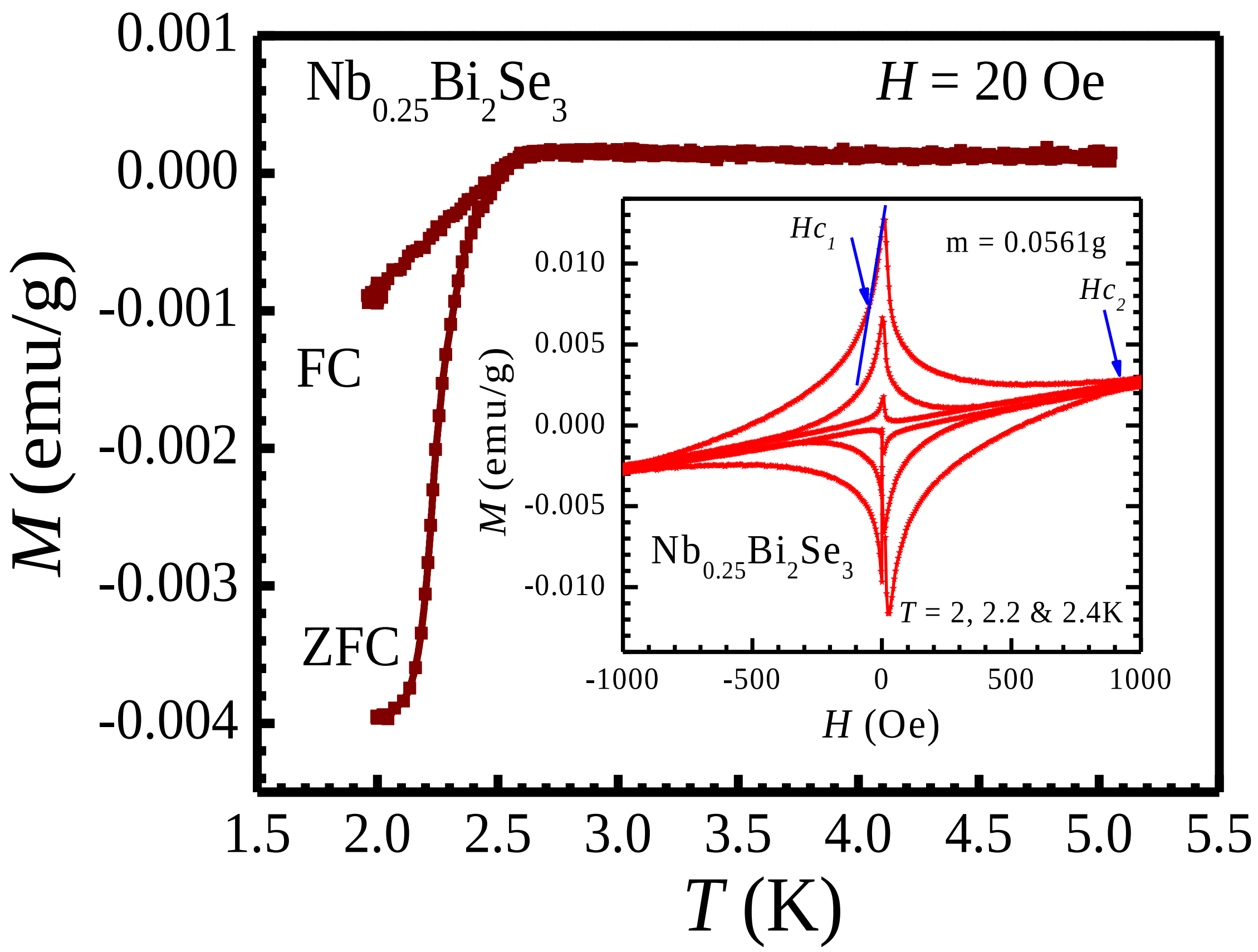